\documentclass[showpacs,pra,aps]{revtex4}
\usepackage{graphics}
\def\Tc{{T^{(0)}_c}}
\def\bfr{{\bf r}}
\def\bfp{{\bf p}}
\def\be{\begin{equation}}
\def\ee{\end{equation}}
\def\bea{\begin{eqnarray}}
\def\eea{\end{eqnarray}}

\begin{document}
\title{Off-diagonal 
long-range order in a harmonically confined two-dimensional Bose gas}
\author{Brandon P. van Zyl}
\affiliation{Department of Physics and Astronomy,
McMaster University, Hamilton,
Ontario, Canada, L8S~4M1}
\begin{abstract}
We investigate the presence of off-diagonal long-range order in a
harmonically confined two-dimensional Bose gas.
In the noninteracting case, an analytical calculation of the
the finite-temperature one-particle density martix 
provides an exact description of the
spatial correlations known to be associated with the existence of a 
Bose-Einstein condensate below the transition temperature $T^{(0)}_c$.
We treat the effects of repulsive interactions within the semiclassical 
Hartree-Fock-Bogliubov approximation
and find that even though the system remains in the same 
{\em uncondensed phase} 
for all $T \geq 0$, there appears to be a revival of off-diagonal long-range 
order for temperatures $T < T^{(0)}_c$.
We suggest that this reentrant order is related to a phase transition 
in the system which {\em is not} the BEC state.
\end{abstract}
\maketitle

\section{Introduction}
Recent advances in the controlled fabrication and manipulation of 
ultra-low temperature trapped Bose gases \cite{gorlitz,bloch} has finally
made it possible to experimentally investigate the existence of
off-diagonal long-range order (ODLRO), i.e., Bose-Einstein condensation (BEC),
in dimensions lower than three.
This remarkable experimental achievment has rekindled an interest in the
classic theoretical problem concerning the existence of BEC in 
two-dimensions (2D).
Although it is well known that finite-temperature BEC can never
occur in a homogeneous 2D Bose gas
\cite{hohenberg},
the ideal trapped Bose gas can undergo BEC below some 
critical temperature $T_c^{(0)}$. 
In the case of a 2D isotropic harmonic trap with confining frequency $\omega_0$
(unless stated otherwise, 
we always implicitly assume this confinement geometry),
Bagnato and Kleppner have shown that
$1/\beta^{(0)}_c = \hbar\omega_0\sqrt{6N/\pi^2}$ \cite{bagnato}, where
$\beta = 1/(k_B T)$; 
Shevchenko later extened this result to include more general power law 
potentials \cite{shevchenko1}.  In the thermodnyamic limit, viz., 
$N \rightarrow \infty$, 
$\omega_0 \rightarrow 0$, such that $N^{1/2}\omega_0 = {\rm constant}$, BEC
is also known to theoretically occur with the critical temperature remaining 
unchanged.
Note that this is not the usual thermodynamic limit, which in 2D would demand
that $N\omega_0 = {\rm constant}$, resulting in no BEC
(see also Ref.~\cite{lieb} for an alternative thermodynamic limit in the
case of an interacting Bose gas).

If repulsive interactions between bosons are included,
(we only consider repulsive interactions in this paper),
then a definitive answer to the question of BEC in the trapped 2D gas 
is not so clear.
In fact, there is no {\it a priori} reason for the existence of
BEC in a confined two or three-dimensional system---BEC is a
purely kinematical phenomenon in the sense that even the ideal
Bose gas can undergo the phase transition.  In contrast,
a transition to the superfluid state is wholey dynamical and
{\em cannot} occur without interactions.
Therefore, one could sensibly assume that interactions prevent
the spatial redistribution of atoms required to achieve Bose gas degeneracy,
and therefore destroy the BEC phase transition altogether.

It is only recently that a rigorous proof has been given to show that there
is no finite temperature BEC in the interacting trapped 2D Bose gas in the
thermodynamic limit \cite{mullin1}. 
Of course, in actual
experiments, the theormodyamic limit is never fully realized, and it becomes
necessary to consider systems consisting of a {\em finite} number of atoms
confined in a trap.  In this regard,
Lieb and Seiringer \cite{lieb}
have proved that the finite, interacting trapped Bose gas does have 100\% BEC
at $T=0$.
To date however, no irrefutable case has been made 
(experimentally or theoretically) in support of finite temperature BEC in an 
interacting finite 2D system.

In a recent paper,
Petrov {\em et al.}~\cite{petrov}
have suggested that that there are in fact two BEC regimes for finite, trapped,
(quasi)-2D systems.  The general argument supporting their claim hinges on 
an approximate
analytical calculation of the one-particle 
density martix from which information about the phase correlations of the
condensate can be obtained.
For large particle number and sufficiently low temperatures,
Petrov {\em et al.} deduce that the  phase fluctuations 
are governed by $\sim T\ln(N)$ \cite{petrov}.
Thus, at temperatures well below $T_c$, fluctuations in the phase are
supressed and one has a ``true condensate'' whereas for 
$T < T_c$,
the phase fluctuations are enhanced and one has a ``quasi-condensate''.
Similar results have recently been obtained by Bogliubov
{\em et al.,} in strictly two-dimenions~\cite{bogliubov}.

However, Bhaduri {\em et al.}~\cite{bhaduri} 
have suggested that there is no BEC
phase transition in a finite, interacting 2D Bose gas 
(see also Refs.~\cite{shevchenko2,vanzyl,mullin2} for similar conclusions).
In their investigation, a semiclassical approximation to the 
Hartree-Fock-Bogliubov (HFB) mean-field theory was used
to investigate the thermodynamics of the trapped 2D Bose gas.  Although the
semiclassical HFB approach could not shed any light on the question 
of phase fluctuations, it was shown that the HFB equations could be solved 
self-consistently all the way down to $T=0$ provided
one {\em does not} invoke the presence of a condensate.
This finding seems to indicate that that if no
{\it a priori} assumption of a condensate is made (as it should be), 
the system remains in the same {\em uncondensed phase} at all 
temperatures. (Note that applying the same HFB approach in 3D requires a 
condensate in order to obtain self-consistent solutions below 
$T^{(0)}_c$.) In spite of the absence of a condensate order-parameter however, 
the density distribution for the uncondensed phase was found to clearly
exhibit a large spatial accumulation of atoms near the center of the trap 
for $T <T^{(0)}_c$,
similar to what is found in systems with a condensate.
This suggests that there may be some sort of ``revival'' of ODLRO
in the system, even though it does not lead to the formation of a condensate.
Since the off-diagonal one-particle density matrix can provide information 
about phase correlations and ODLRO in the system, an investigation of 
the off-diagonal density martix within the HFB approximation needs to be 
performed.  This is indeed the central motivation for the present work.

The rest of our paper is organized as follows.  In Sec.~\ref{noninteracting} we
investigte the noninteracting 2D gas in a harmonic trap where
an exact expression for the finite temperature correlation function can be 
obtained.  In Sec.~\ref{interacting}, we calculate
the correlation function for the interacting system within the semiclassical 
HFB approximation and investigate its temperature dependence both above and 
below $T^{(0)}_c$.  Finally, in Sec.~\ref{conclusions}, we present
our concluding remarks.

\section{Noninteracting gas}
\label{noninteracting}
In this section, we wish to pay special
attention to the off-diagonal, one-particle density matrix 
$\rho^{(0)}(\bfr,{\bf r}';\beta)$ of the noninteracting, confined 2D Bose
gas.  Our motivation lies in the fact that ODLRO in the one-particle 
density matrix and BEC are interrelated \cite{penrose} (see below for details). 
For the isotropic HO, $\rho^{(0)}(\bfr,{\bf r}';\beta)$ can be calculated 
analytically, and we have an exact result from which 
we can quantitatively test the 
validity of the semiclassical approximation.  We will be particularly 
interested how well the seimclassical approximation describes the phase
correlations of the gas below the BEC critical temperature $\Tc$.  
To the best of our
knowledge, this is the first such study in two-dimensions.

By definition, the off-diagonal one-particle density matrix is given 
by \cite{fetter}
\be
\rho^{(0)}(\bfr,{\bf r}';\beta) = \sum_j \psi^{(0)}_j(\bfr)
\psi^{(0)}_j(\bfr')^{\dagger} n^{(0)}_j~,
\label{green}
\ee
where $\psi^{(0)}_j(\bfr)$ are exact eigenfunctions of the $d$-dimensional 
harmonic
oscillator (HO) and $n^{(0)}_j$ is the Bose distribution function.  
It is most convenient to evaluate (\ref{green}) for the
1D case, and then use the separability of the HO potential to obtain the 
general result.  Explicitly, we have
\be
\rho^{(0)}(x,x';\beta) = \frac{1}{\sqrt{\pi}}\sum_{n=0}^{\infty}e^{-\frac{1}{2}
(x^2 + x'^2)}\frac{H_n(x)H_n(x')}{2^nn!}\frac{1}
{e^{\beta(\varepsilon_n-\mu)}-1}~,
\label{green2}
\ee
where $\varepsilon_n$ are the energy levels of the HO, $\mu$ is the 
chemical potential, and we have scaled all energies and lengths by 
$\hbar \omega_0$ and $\ell_0 = \sqrt{\hbar /m\omega_0}$, respectively.
We now set the zero of energy to correspond to the lowest trap level, viz.,
$n=0$, and obtain from (\ref{green2})
\be
\rho^{(0)}(x,x';\beta) = 
\frac{1}{\sqrt{\pi}}\sum_{j=1}^{\infty}e^{j\beta \mu}
\sum_{n=0}^{\infty}e^{-\frac{1}{2}
(x^2 + x'^2)}\frac{H_n(x)H_n(x')}{2^nn!}e^{-j\beta n}~,
\ee
where we have used \cite{ketterle}
\be
\sum_{j=1}^{\infty}e^{j\beta\mu}e^{-j\beta n} = \frac{1}{e^{\beta(\varepsilon_n-\mu)}-1}~.
\ee
The summation over trap levels can be perfomed exactly \cite{morse}, and 
we finally obtain in any dimension
\be
\rho^{(0)}(\bfr,\bfr';\beta) = \sum_{j=1}^{\infty} \frac{e^{j\beta\mu}}
{[\pi(1-e^{-2j\beta})]^{d/2}}
e^{-1/4[|\bfr +\bfr'|^2\tanh(j\beta/2)+|\bfr-\bfr'|^2\coth(j\beta/2)]}~.
\ee
Putting $d=3$ in the above equation reproduces the result obtained by
Barnett {\em et al}.,~\cite{barnett}.
The chemical potential is obtained by enforcing that the total particle 
number $N$, remain fixed, viz.,
\be
N = \sum_{j=1}^{\infty}\frac{e^{j\beta\mu}}{(1-e^{-j\beta})^{d}}~.
\ee
\begin{figure}
\resizebox{5in}{5in}{\includegraphics{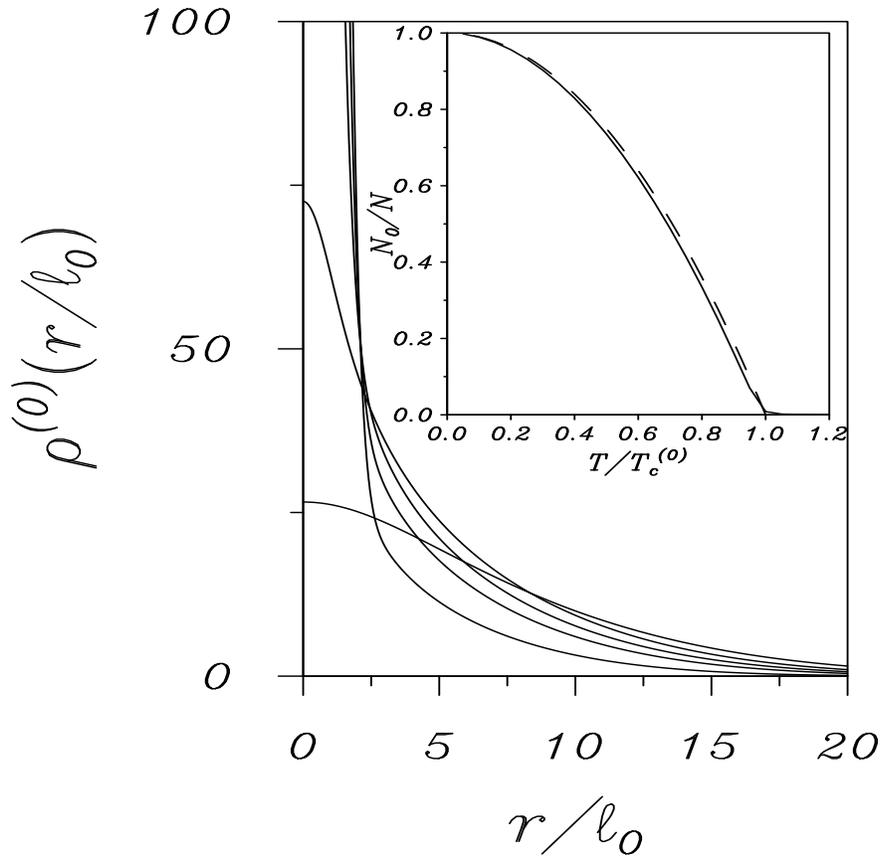}}
\caption{Exact quantum-mechanical densities for temperatures
(starting from the lowest curve) $T/\Tc = 1.2, 1.0, 0.9,0.8,0.6$.
The figure inset shows the fractional occupancy $N_0/N$ of the lowest
oscillator level for $N=10^4$ atoms (solid curve) and the thermodyanmic
limit (dashed line) as given by Eq.~(7).}
\end{figure}
The diagonal part of the one-particle density matrix yields the exact
quantum-mechanical finite 
temperature density, $\rho^{(0)}(\bfr)$, of the trapped gas. 
The densities for various temperatures are shown in Fig.~1 for
$d=2$ and $N=10^4$ atoms.  Figure 1 illustrates that even though we are not
in the thermodynamic limit, the single-particle density distribution
exhibits an enormous
accumulation of atoms near the center of the trap for $T < T_c^{(0)}$. 
This type of behaviour is consistent with the confined gas
having undergone a phase transition to the BEC state. 
The usual approach for confirming that this phase transition is indeed BEC 
is to examine the temperature dependence of the fractional occupancy 
of the lowest oscillator level, $N_0/N$.
The inset to Fig.~1 shows that below $T/T_c^{(0)}\approx 1$, 
the ground state of the system becomes macroscopically occupied, with already
$50 \%$ of the atoms in the lowest level at $T/\Tc \approx 0.7$.
Remarkably, even at $N=10^4$ atoms, the fractional
occupancy closely follows the quadratic temperature dependence
obtained strictly in the thermodynamic limit (dashed line in figure inset), 
viz.,
\be
\frac{N_0}{N} = 1 - \left(\frac{T}{T_c^{(0)}}\right)^{2}~.
\ee
Although $N_0/N$ gives us a strong indication that the finite system has a BEC
transition at $T/T_c^{(0)}\approx 1$, it does not provide us with a
definitive test for the presence of ODLRO in the system.

It is well known since the early work of Penrose and Onsager \cite{penrose}
that for a {\em homogeneous} system, the
phenomenon of BEC is intimately related to the presence of ODLRO in the
off-diagonal one-particle density matrix, namely,
\be
\lim_{|\bfr-\bfr'|\rightarrow \infty}\rho^{(0)}(\bfr,\bfr';\beta) \neq 0~.
\ee
In fact, the low temperature behaviour of $\rho^{(0)}(\bfr,\bfr')$
is directly related to the phase fluctuations of the condensate \cite{penrose}
and the non-zero limiting value of Eq.~(8)
yields the  value of the order parameter (i.e., condensate fraction) in the
system.
For finite inhomogeneous systems however, 
the above criterion has to be modified to account for the
fact that ODLRO cannot extend beyond the physical size of the system.
This is accomplished by defining a
normalized correlation function (sometimes called the first-order coherence
function) by
\be
g^{(0)}(\bfr,\bfr';\beta) = \frac{\rho^{(0)}(\bfr,\bfr';\beta)}
{\sqrt{\rho^{(0)}(\bfr,\bfr;\beta)}\sqrt{\rho^{(0)}(\bfr',\bfr';\beta)}}~,
\ee
such that
\be
\lim_{|\bfr-\bfr'|\rightarrow \xi}g^{(0)}(\bfr,\bfr';\beta)\neq 0~. 
\ee
The above correlation function describes a local measure of coherence
in the system.  For example, the ability of two BEC's, intially 
separated by $|\bfr - \bfr'|$, to form an interference pattern is quantified
by $g^{(0)}(\bfr,\bfr')$ \cite{hecht}.  
In this sense, the correlation function also characterizes
local fluctuations of the phase of the condensate order-parameter.
The quantity $\xi$ is the large-distance length-scale over which the
ODLRO of the finite system is manifested.  At zero temperature,
$\xi = \ell_0$, whereas for $T\neq 0$, $\xi$ is taken to be the spatial
extent of the single-particle density $\rho^{(0)}(\bfr)$.  To simplify our
calculations, we fix one of the coordinate, $\bfr'$ to be at the center of the
trap.
Note that in Eq.~(9), dividing the off-diagonal density matrix by the
square root of the single-particle densities removes any effects of the 
local-density of particles on the correlation function, and gives 
$g^{(0)}(\bfr,\bfr;\beta) = 1$ in the case of perfect coherence.

\begin{figure}
\resizebox{5in}{5in}{\includegraphics{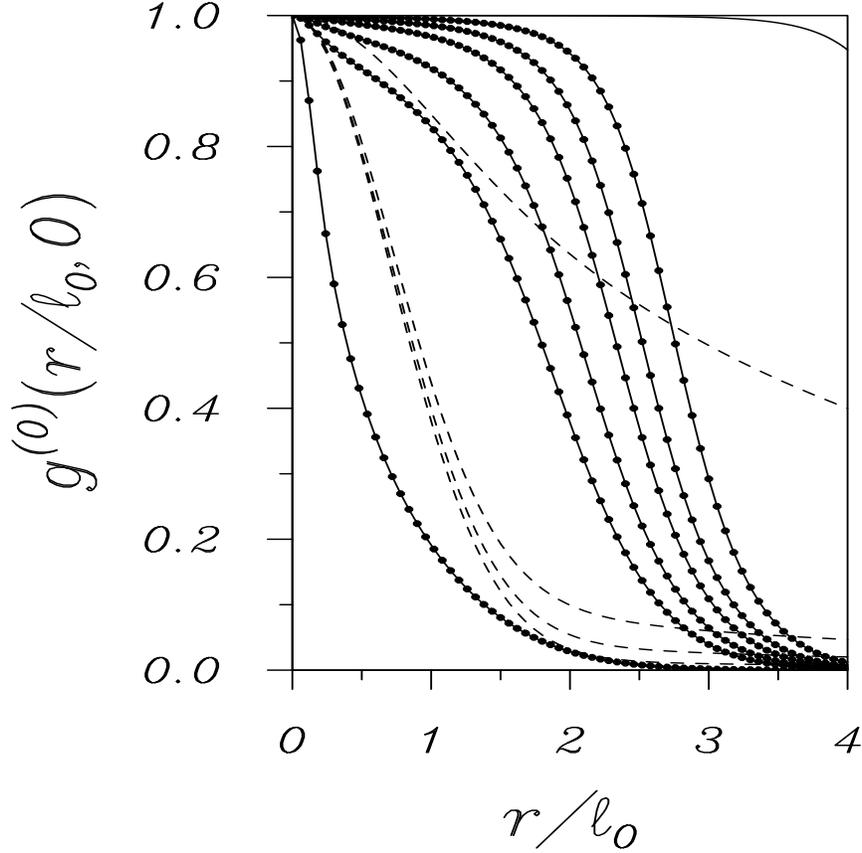}}
\caption{Exact off-diagonal correlation function (solid curves) for
temperatures (from left to right) $T/\Tc = 1.0,0.9,0.8,0.4,0.2,0.01$.
The dashed lines correspond to the scaled exact quantum mechanical
densities at temperatures (from right to left) $T/\Tc = 1.0,0.9,0.8,0.6$
The filled circles are the semiclassical approximation to the
off-diagonal density matrix as given by Eq.~(16).
}
\end{figure}

The solid curves in Fig.~2 depict the correlation function $g^{(0)}(\bfr,0)$
for $d=2$ and $N=10^4$ at various temperatures.
From left to right, we have $T/\Tc = 1.0,0.9,0.8,0.6,0.4.0.2,0.01$.
Above the critical temperature,
the atoms remain in the uncondensed phase and the system
shows no ODLRO.  However, as the gas is cooled below the transition
temperature, we see an abrupt jump in the large-distance decay of the
spatial correlations.  A direct application of Eq.~(10) yields
condensate fractions of 
$N_0/N = 0.02,0.47,0.71,0.84, 0.93$, 
for temperatures $T/T_c^{(0)} = 1.0, 0.8, 0.6, 0.4, 0.2$, respectively.
Thus, below $\Tc$, the finite noninteracting 2D Bose gas contains a
non-zero order-parameter (i.e., a BEC).

It also proves very
useful to compare the spatial extent of the single-particle density,
$\rho^{(0)}(\bfr)$, to the spatial decay of the correlation function
$g^{(0)}(\bfr,0)$.  This information is also presented in Fig.~2
where the exact quantum mechanical densities are represented by dashed lines.
These densities have been scaled by their value at $\bfr = 0$ to allow
for a comparison of the length scales between $\rho^{(0)}(\bfr)$ and
$g^{(0)}(\bfr,0)$.
From right to left, the dashed lines correspond to densities at 
$T/\Tc = 1.0, 0.9,0.8,0.6$.  Densities at lower temperatures are almost
indistinguishable on the scale of the figure.
We note
that at low temperatures, 
the correlation function is practically a constant $g^{(0)}(\bfr,0) = 1$
over the entire region for which the density (i.e., physical size of
the system) of the gas is nonzero.  
This means that phase fluctuations are very small, and the density 
distribution corresponds to that of a
``true condensate''.  At higher temperatures (i.e., $0.9 < T/\Tc < 1.0$)
the single-paricle density can become comparable to that of $g^{(0)}(\bfr,0)$,
and the gas can be said to be a ``quasi-condensate'' in the sense that
the phase correlations do not extend to all of the atoms in the
trap.  For $T/\Tc > 1.0$, the spatial extent of the single-particle density
greatly exceeds characteristic length of the correlations, and system
is in the normal state.

Having obtained the exact expression for $\rho^{(0)}(\bfr,\bfr';\beta)$,
we now wish to examine its semiclassical approximation, namely, its behaviour
when $k_BT \gg \hbar\omega_0$.  It is easy to show that Eq~(5) reduces to
\begin{eqnarray}
\rho^{(0)}(\bfr,\bfr')&=& \frac{1}{(2\pi\beta)^{d/2}}\sum_{j=1}^{\infty}
\frac{e^{j\beta\mu}}{j^{d/2}}\exp\left[
-\beta [V(\bfr)+V(\bfr')]/2\right]^j\exp\left[-|\bfr-\bfr'|^2)/2\beta\right]^{1/j}\\
&=& \frac{1}{(2\pi\beta)^{d/2}}g_{d/2}(\exp\left[
\beta(\mu- [V(\bfr)+V(\bfr')]/2)\right],\exp\left[-|\bfr-\bfr'|^2)/2\beta\right])~,
\end{eqnarray}
where $V(\bfr) = 1/2 r^2$, and as in Ref.~\cite{nara} we have introduced 
the generalized Bose function
\be
g_{\alpha}(x,y) = \sum_{j=1}^{\infty}\frac{x^j y^{1/j}}{j^{\alpha}}~.
\ee
For $\bfr = \bfr'$ and $d=2$, Eq.~(12) becomes
\be
\rho^{(0)}(\bfr) = \frac{1}{(2\pi)^2} \int\frac{d^2\bfp}
{\left[\exp \left[\left(\frac{p^2}{2} + \frac{1}{2}r^2 + -
\mu\right)\beta\right] -1 \right] }~,
\ee
with the normalization condition
\be
N = \int~\rho^{(0)}(\bfr)~d^2\bfr~.
\ee

If one attempts to solve Eqs.~(14,15) 
for all $T$, it is readily shown that there exists a temperature $T^{\star}$ 
below which the
equations can no longer be satisfied~\cite{bhaduri,vanzyl}.  In fact, the
temperature $T^{\star}$ is identical to the the critical temperature
$T_c^{(0)}$ obtained by Bagnato and Kleppner~\cite{bagnato}.  
Thus, in order to obtain solutions below
$T_c^{(0)}$, a condensate order parameter is required.  We introduce the
macroscopic occupation of the ground state in the one-particle density matrix
according to the {\it ansatz}
\be
\rho^{(0)}(\bfr,\bfr') = N_0 \psi^{(0)}_0(\bfr) \psi^{(0)}_0(\bfr') +
\frac{1}{(2\pi\beta)^{d/2}}g_{d/2}(\exp\left[
\beta(\mu- [V(\bfr)+V(\bfr')]/2)\right],\exp\left[-|\bfr-\bfr'|^2/
2\beta\right])~,
\ee
where $ \psi^{(0)}_0(\bfr)$ is the zero temperature
ground-state eigenfunction of the HO and
\be
N_0 = \frac{e^{\beta\mu}}{1-e^{\beta\mu}}~.
\ee
The 2D density is then given by Eq.~(16) with $\bfr = \bfr'$,
\be
\rho^{(0)}(\bfr) = N_0|\psi^{(0)}_0(\bfr)|^2 + 
\frac{1}{(2\pi)^2} \int\frac{d^2\bfp}
{\left[\exp \left[\left(\frac{p^2}{2} + \frac{1}{2}r^2 + -
\mu\right)\beta\right] -1 \right] }~.
\ee

The semiclassical densities obtained from Eq.~(18) are indistinguishable from
the exact densities obtained from Eq.~(5).
The semiclassical correlation function, Eq.~(16), at various temperatures is 
represented in Fig.~2 as filled circles.  
The superb agreement between the exact (solid curves) and semiclassical
results shows that the
semiclassical approximation is valid well below the critical temperature 
provided one takes into account the macroscopic occupation of the ground
state separately. If the atoms had remained in the uncondensed phase, 
Eqs.~(14,15) would have been sufficient to describe the gas for all $T\geq 0$.

\section{Interacting gas}
\label{interacting}

The thermodynamic properties of the interacting Bose gas are most commonly 
described by the self-consistent mean-field HFB equations.
In most calculations (see Ref.~\cite{giorgini,dalfovo} 
and references therein for
details), the semiclassical approximation is applied to the discrete set 
of coupled Bogliubov equations, leading to the considerably simplified 
semiclassical HFB theory.
The semiclassical HFB approximation is well-known to be an excellent
tool for describing the interacting 3D Bose gas \cite{giorgini,dalfovo}.
The 2D version of the semiclassical HFB theory is formally identical to the
3D version and has already been examined by Mullin
\cite{mullin3}.  We present only the essentials of the model here for
completeness. 

In the semiclassical HFB approach, the condensed and uncondensed atoms are 
described by a set of coupled, self-consistent equations.
The macroscopic condensate wavefunction, $\psi_0(\bfr)$, is 
characterized by the finite-temperature Gross-Pitaevskii equation
which is given by
\be
\left[-\frac{1}{2}\nabla^2 + \frac{1}{2}r^2 - \mu + 2\gamma\rho(\bfr) -
\gamma \rho_0(\bfr)\right]\psi_0(\bfr) = 0~,
\ee
where $\gamma$ is the coupling constant associated with the assumed
repulsive zero-range two-body pseudo-potential. 
The total density, $\rho(\bfr)$, is made up of a condensate 
$\rho_0(\bfr) \equiv |\psi_0(\bfr)|^2$, and a noncondensate density,
\be
\rho_T(\bfr) =  \int~\left\{
\left[u^2(\bfp,\bfr) + v^2(\bfp,\bfr)\right] f(\bfp,\bfr) +
v^2(\bfp,\bfr)\right\}~d^2\bfp~,
\ee
where
\be
f(\bfp,\bfr) = \frac{1}{e^{\beta\varepsilon}-1}~,
\ee
\be
u^2(\bfr,\bfp) = \frac{\Lambda +\varepsilon}{2\varepsilon}~,
\ee
\be
v^2(\bfr,\bfp) = \frac{\Lambda -\varepsilon}{2\varepsilon}~,
\ee
\be
\Lambda = \frac{p^2}{2} + \frac{1}{2}r^2 - \mu + 2\gamma\rho(\bfr)~,
\ee
and
\be
\varepsilon(\bfp,\bfr) = \sqrt{\Lambda^2 - (\gamma\rho_0(\bfr))^2}~.
\ee
The total number of particles satisfies 
\be
N = \int~[\rho_0(\bfr) + \rho_T(\bfr)]~d^2\bfr.
\ee
In the absence of a condensate, it is easy to show that
the above set of HFB equations reduce to
\be
\rho_T(\bfr) = \frac{1}{2\pi}\int\frac{d^2\bfp}
{\left[\exp \left[\left(\frac{p^2}{2} + \frac{1}{2}r^2 + 2\gamma\rho_T(\bfr) -
\mu\right)\beta\right] -1 \right] }~,
\ee
with
\be
N = \int~\rho_T(\bfr)~d^2\bfr.
\ee
\begin{figure}
\resizebox{5in}{5in}{\includegraphics{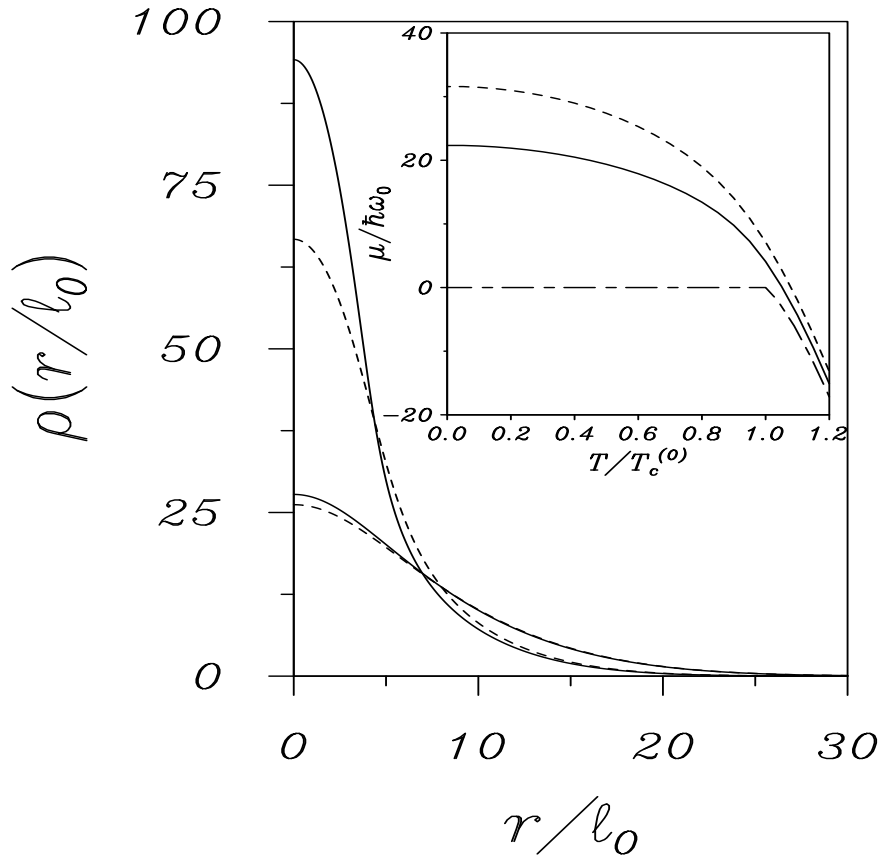}}
\caption{Semiclassical densities for $N=10^4$ atoms with
$\gamma = 0.16$ (solid curves) and $\gamma=0.31$ (dashed curves).
The lowest lying curves correspond to $T/\Tc = 1.1$ and the higher lying
curves to $T/\Tc = 0.8$.  The figure inset shows the chemical potential
as a function of temperature.  The long-dashed-short-dashed curve
corresponds to $\gamma = 0$.  The value $\gamma = 0.31$ is consistent
with an effective 2D interaction strength for Rb$^{87}$ \cite{bhaduri}.}
\end{figure}
Equations (27,28) are equivalent to the so-called self-consistent
Thomas-Fermi approximation used in Refs.~\cite{bhaduri,vanzyl}.
Putting $\gamma=0$ in Eq.~(27) yields an expression that is identical
to the semiclassical approximation of the exact diagonal density matrix, 
viz., Eq.~(14), found in the previous section;
recall from Sec.~\ref{noninteracting} that below $\Tc$, Eqs.~(27,29)
cannot be solved self-consistently.  
However, if one puts any finite $\gamma > 0$ in Eq.~(27), 
self-consistent solutions to the HFB equations can be obtained all the way 
down to  $T  = 0$~\cite{bhaduri,vanzyl,mullin2}.  
To illustrate this, we show in Fig.~3 the density and chemical potential
(figure inset) of the gas for $\gamma = 0$ (long-dash-short-dashed line),
$\gamma = 0.16$ (solid line), and $\gamma = 0.31$ (dashed line).
In contrast to the noninteracting 2D gas 
(where a condensate order-parameter has to be introduced below $\Tc$), 
the interacting system remains in the uncondensed phase for all $T\geq 0$.   

Although no condensation occurs, it is interesting to compare the densities
of the noninteracting (Fig.~1) and interacting gases (Fig.~3)
just above and below $\Tc$.  The essential point to be made from this 
comparison is that in both cases, there is an
increase in the density of atoms near the center of the trap for $T < \Tc$.
While the noninteracting density shows a much larger enhancement
at $\bfr = 0$, the interacting gas also displays a similar (albeit diminished)
behaviour.
Since an increase in the density of atoms near the center of the trap 
is associated with ODLRO in the noninteracting gas, it is worthwhile 
investigating the off-diagonal
density matrix for the interacting 2D gas within the semiclassical
HFB approximation.  

The 2D semiclassical HFB off-diagonal density matrix can be obtained from
the leading order term in the Wigner-Kirkwood semiclassical expansion of the 
quantum mechanical off-diagnonal density matrix \cite{brack}.   The
noncondensed particles are treated as bosons in an effective
potential, and we obtain
\be
\rho(\bfr,\bfr';\beta) = \frac{1}{2\pi\beta}g_1
(\exp\left[
\beta(\mu- [V_{\rm eff}(\bfr)+V_{\rm eff}(\bfr')]/2)\right],\exp\left[-|\bfr-\bfr'|^2/
2\beta\right])~,
\ee
where
\be
V_{\rm eff} (\bfr) = \frac{1}{2}r^2 + 2 \gamma \rho(\bfr)~.
\ee
Since there is no condensate for $\gamma >0$, we {\em have not} included the 
first term in Eq.~(16).  Of course, for $\bfr=\bfr'$, Eq.~(29) reduces to
Eq.~(27).  

As in Sec.~\ref{noninteracting}, the normalized correlation function for the
finite sized system is given by
\be
g(\bfr,0;\beta) = \frac{\rho(\bfr,0;\beta)}{\sqrt{\rho(\bfr)}\sqrt{\rho(0)}}~,
\ee
where again, we have fixed $\bfr' = 0$.
The temperature dependence of $g(\bfr,0;\beta)$ is
shown in Figs.~4, 5 for coupling strengths $\gamma = 0.16, 0.31$, respectivley.  
\begin{figure}
\resizebox{5in}{5in}{\includegraphics{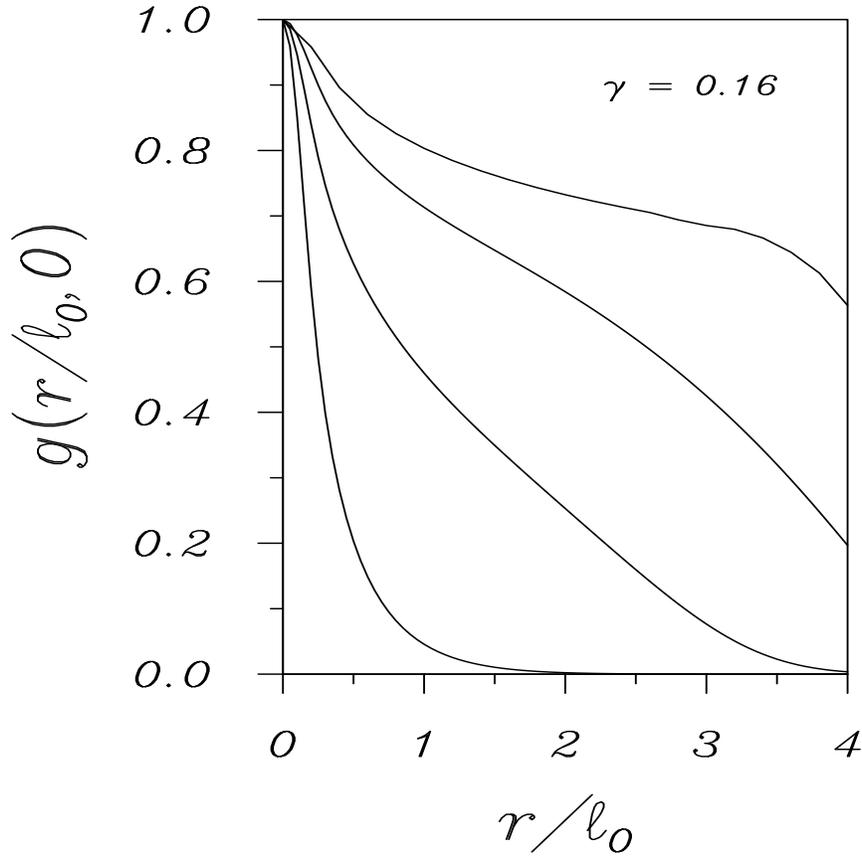}}
\caption{The semiclassical HFB approximation to the correlation function with
$\gamma = 0.16$.  The curves correspond to (from left to right)
$T/\Tc = 1.0,0.8,0.6,0.4$}
\end{figure} 
\begin{figure} 
\resizebox{5in}{5in}{\includegraphics{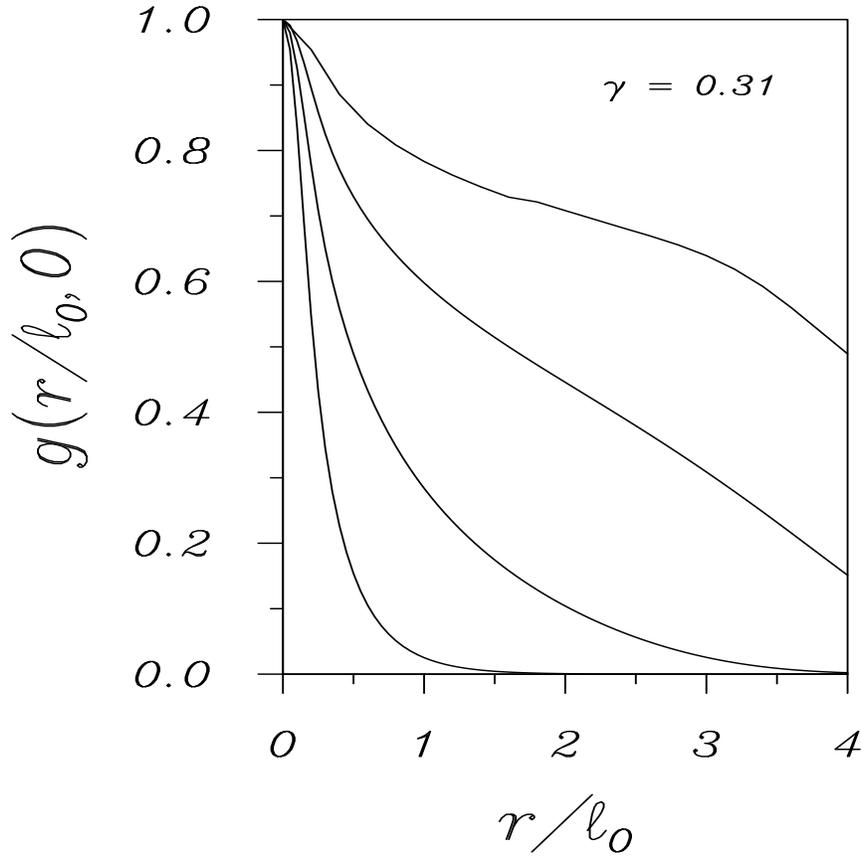}}
\caption{As in Fig.~4 but for $\gamma = 0.31$.}
\end{figure}
Above $\Tc$, the interacting and noninteracting 
(see Fig.~2) systems have similar correlations.  
However, as we cool the gas to temperatures just below $\Tc$, there is an
drastic change in the behaviour of the interacting 
off-diagonal density matrix, which to the best of our knowledge, has never
been noticed before.
In particluar, we observe that the decay of the
correlations in the interacting system 
are qualitatively different from the noninteracting gas.
For the interacting gas, values of $r \gtrsim \ell_0$ lead to the tail of 
the correlation function decreasing 
slowler than the noninteracting case.   
In contrast, for $r \lesssim \ell_0$, the interacting correlation
function has a much faster spatial decay than the noninteracting system.
We have already discussed the fact that the near constancy of 
the noninteracting correlation function, $g^{(0)}(\bfr,\bfr')$, over the 
size of the sample ($r \approx \ell_0$)
is indicative of negligible phase fluctuations in the condensate 
\cite{penrose}.  
Thus, the re-appearance of ODLRO in $g(\bfr,\bfr')$ below $\Tc$
suggests that there may be larger fluctuations in the phase as compared to the 
noninteracting gas.
If we apply Eq.~(10) to $g(\bfr,\bfr')$ (with $\gamma = 0.16$), we obtain 
$0.002, 0.25, 0.58, 0.70$ at $T/\Tc = 1.0, 0.8, 0.6, 0.4$ respectively, for  
the asymptotic behaviour of the correlation fucntion.  Even though 
we may not be able to interpret these numbers as condensate fractions, it is 
tempting to conjecture that they represent a nonzero value of an 
order-parameter in the system.

We can be more quantitative about the nature of the correlations 
displayed by $g(\bfr,\bfr')$ below $\Tc$ as follows.  In the interacting
{\em homogeneous} 2D Bose gas, Popov \cite{popov} has already calculated
the low temperature  correlation function:
\be
g(\bfr,\bfr';\beta) \approx R^{-\alpha}~,
\ee
where 
\be
R \equiv \frac{|\bfr-\bfr '|}{\beta\sqrt{\gamma\rho_s}},~~~~~~
\alpha = \frac{1}{2\pi\beta \rho_s}~,
\ee
and $\rho_s$ is the so-called superfluid density.  The
{\em algebraic} ODLRO exhibited by $g(\bfr,\bfr')$
implies that at low temperature,
correlations can extend over macroscopic (i.e., size of the system)
distances, although phase fluctutations will be nonzero.  This is 
reminiscent of the quasi-condensate described in Ref.~\cite{petrov},
and it is only identically at $T=0$ that the system will attain a ``true
condensate''.

If the system is finite (i.e., in a 2DHO potential), then Bogliubov
{\em et al.}~\cite{bogliubov}, have shown that 
\be
R \equiv \frac{|\bfr-\bfr '|}{\beta\sqrt{\gamma\rho(0)}},~~~~~~
\alpha = \frac{1}{2\pi\beta \rho({\bf S})}~,
\ee
where ${\bf S} = (\bfr+\bfr')/2$.
Eq.~(34) is formally identical to (33) except for the slowly varying
factor $\rho({\bf S})$ which is the inhomogeneous boson density.
This additional spatial dependence results in a deviation from the
strict power law dependence of the uniform gas.
We have found that at low temperatures, 
the power law dependence in Eq.~(34) agrees well with
the observed numerical behaviour of $g(\bfr,\bfr';\beta)$.
This suggests that at low temperatures, the interacting system exhibits
algebraic ODLRO similar to that of the superfluid state in the uniform
Bose gas. At exactly $T=0$, the gas will have a true condensate.

\section{Conclusions}
We have investigated the finite-temperature correlations in both the ideal 
and interacting trapped, 2D Bose gas.  Our study of the noninteracting
gas provided an exact framework from which we could compare the validity of
the semiclassical approximation over a wide temperature range.
Our results indicate that the semiclassical approach works extremetly well,
even below the critical temperature, provided the macroscopic occupations of
the ground state is treated separately.  A careful examination of the
correlation function reveals that, as in Ref.~\cite{petrov},
there are two BEC regimes for the noninteracting gas.  Namely, for temperatures 
$0.9 < T/\Tc < 1.0$, the system is a quasi-condensate in the sense that the
phase correlations are comparable to the spatial extent of the single-particle
density.  On the other hand, when the temperature is lowered to
$T/\Tc \lesssim 0.9$, phase fluctuations become negligible, and the gas
is a true condensate.  It is straightforward to apply our results 
to study the finite $N$ dependence on the coherence properties
of trapped gases in dimensions $d=1,2,3$.
(see e.g., Ref.~\cite{barnett} for work relating to 3D).

In the interacting gas, we found that even though the system does not contain
a condensate, there is a revival of ODLRO for $T/\Tc \lesssim 1$.
The low temperature behaviour of the interacting correlation function
was found to be consistent with an algebraic (i.e., power law) 
decay in the phase correlations, similar to what is found in the uniform 2D
superfluid Bose gas.  Whether or not this re-entrant ODLRO is to the
superfluid state is presently being investigated.

Finally, it is interesting to note that Margo and Ceperely 
\cite{margo} have found conclusions analgous to ours 
for the uniform 2D charged Bose fluid (CBF) with $\ln (r)$ interactions.  
Specifically, they found that even though the 2D CBF
{\em does not} undergo a condensation, the noncondensed fluid exhibits a
power-law decay of the one-particle density matrix similar to what is found
in the present work.  It is not immediately clear what relevance their 
findings may have on our results for the interacting, 
inhomogeneous 2D Bose gas.

\label{conclusions}

\begin{acknowledgments}
It is a pleasure to thank Drs.~Wonkee Kim, Rajat Bhaduri and J. P. Carbotte 
for illuminating discussions.  
This work was supported in part by the National Sciences and the Engineering
Research Council of Canada (NSERC) and by the Canadian Institute for Advanced
Research (CIAR).
\end{acknowledgments}

\bibliographystyle{pra}

\end{document}